\documentstyle[psfig,times]{mn}
\def\H0{{\it H}$_0$}

\def\q0{{\it q}$_0$}

\def\psqcm{cm$^{-2}$}

\title[PMN~J0525-3343 ] {PMN~J0525-3343: soft X--ray spectral
flattening in a blazar at z=4.4} 
\author[A.~C. Fabian et al.]  
{\parbox[]{6.5in} 
{A.~C. Fabian,$^1$ A. Celotti,$^2$ K. Iwasawa,$^1$
R.~G. McMahon,$^1$ C.~L. Carilli,$^3$  W.~N. Brandt,$^4$ 
G.~Ghisellini$^5$ and I.~M. Hook$^6$}\\ 
\\
$^1$Institute of Astronomy, Madingley Road, Cambridge CB3 0HA\\ 
$^2$S.I.S.S.A., via Beirut 2-4, I-34014 Trieste, Italy\\ 
$^3$ NRAO, P.O. Box 0, Socorro NM, 87801, USA\\ 
$^4$Department of Astronomy and Astrophysics, The Pennsylvania
State University, 525 Davey Lab, University Park, PA 16802, USA\\
$^5$Osservatorio Astronomico di Brera-Merate, via Bianchi 46, 
I-23807 Merate (LC), Italy \\ 
$^6$ IFA, Royal Observatory, Blackford Hill, Edinburgh EH9 3HJ \\
} 

\date{}

\begin{document}

\maketitle

\begin{abstract}
We report optical, radio and X--ray observations of a new distant
blazar, PMN~J0525-3343, at a redshift of 4.4. The X-ray spectrum
measured from ASCA and $Beppo$SAX flattens below a few keV, in a
manner similar to the spectra of two other $z>4$ blazars, GB~1428+4217
($z$=4.72) reported by Boller et al and RXJ~1028.6-0844 ($z$=4.28) by
Yuan et al. The spectrum is well fitted by a power-law continuum which
is either absorbed or breaks at a few keV. An intrinsic column density
corresponding to $2\times 10^{23}$H-atoms\ \psqcm at solar abundance is
required by the absorption model. This is however a million times
greater than the neutral hydrogen, or dust, column density implied by
the optical spectrum, which covers the rest-frame UV emission of the
blazar nucleus.  We discuss the problems raised and suggest that,
unless there is intrinsic flattening in the spectral distribution of
the particles/seed photons producing X--rays via inverse Compton
scattering, the most plausible solution is a warm absorber close to
the active nucleus.
\end{abstract}

\begin{keywords}
galaxies: active - galaxies: individual: PMN~J0525-3343 - X-ray: galaxies.
\end{keywords}

\section{Introduction}

Active Galactic Nuclei (AGN) at high redshift are powerful tools with
which to study the evolution of massive black holes and of their young
galaxy hosts. A particularly interesting class is formed by several
high redshift ($z>4$), X-ray bright, radio--loud quasars (Fabian et
al 1997, 1998; Moran \& Helfand 1997; Zickgraf et al 1997; Hook \&
McMahon 1998) which present characteristics typical of blazars. One
such source, GB~1428+4217, is variable in both the X-ray and radio
bands and has a spectral energy distribution which peaks at hard X-ray
energies (Fabian et al 1998).  Recently this object (Boller et al
2000) and RXJ1028.6-0844 (Yuan et al 2000) have been found to show
spectral flattening which is interpreted as due to X-ray absorption,
implying an absorbing column density of $\sim 1.5\times10^{22}$\psqcm
and $\sim 2.1\times 10^{23}$\psqcm, respectively, if intrinsic. We
report here on the discovery and study of a similar object,
PMN~J0525-3443 at $z=4.4$, which shows similar spectral flattening.

Fiore et al (1998) have earlier found from ROSAT data an apparent
systematic decrease with redshift of the spectral slope of the soft
X--ray emission of radio-loud quasars from local to $z\sim 3.9$
objects.  The change in spectra with redshift might be associated with
an increase with redshift in the amount of absorbing (intrinsic or
external) gas. Cappi et al (1997) find evidence for absorption using
ASCA data for a small sample of high-redshift ($1.2<z<3.4$) radio-loud
quasars. Confirmation of a systematic spectral flattening has been
also reported by Reeves \& Turner (2000) for a significant number of
radio--loud sources observed by ASCA up to $z$=4.2.
Understanding the nature and evolution of this spectral change will be
of great importance for our knowledge of gaseous environments of
powerful quasars, their intrinsic properties as well as their
cosmological evolution and the radio--loud quasar/galaxy connection.

The object under study here, PMN J0525-3343, was identified by McMahon
and Hook as part of their program to identify high redshift
radio--loud quasars (Hook \& McMahon 1998).  The quasar was selected
as a flat spectrum (i.e. $\alpha < $ 0.5; $F_{\nu} \propto
\nu^{-\alpha}$) radio source using the 4850 MHz Parkes-MIT-NRAO (PMN)
survey (Wright et al 1996) and 1.4 GHz NRAO VLA Sky Survey (NVSS,
Condon et al 1998).  The 5.0GHz and 20cm fluxes are 210$\pm$18 and
188.7$\pm$5.7 mJy, respectively, giving a spectral index 
$\alpha_{1.49-5.0 GHz} = -0.1$.  The radio sources were
matched against the APM UKST catalogue using the NVSS radio position
and it was identified as a red stellar object with R=18.5 and colour
B--R=2.80. Subsequent optical spectroscopy identified the radio source
as a quasar with z=4.4.  It was confirmed to be an X--ray source with
the ROSAT HRI and we have since observed it with ASCA and $Beppo$SAX.
We present here X-ray fluxes and spectra, together with measurements
of its radio and optical spectra.  PMN~J0525-3443 is the second most
distant object in the X--ray bright, radio--loud class of quasars.

The optical, radio and X--ray data analysis procedure and results are
reported in Sections~2, 3 and 4, respectively, while discussion and
conclusions are presented in Sections~5 and 6.  $H_0$=50 km
s$^{-1}$ Mpc$^{-1}$ and $q_0$=0.5 have been used throughout the paper.

\section{Identification and Optical data}

An X--ray image of the quasar was obtained with the ROSAT HRI on 1998
March 4--6 for 8222~s. Contours of X-ray emission are overlaid on the
POSS optical image in Fig.~1, demonstrating that the X--ray
identification is secure (the approximately 4 arcsec offset is within
the ROSAT positional uncertainties).

An optical spectrum covering the spectral range 4000-9000\AA\, was
obtained on 1998 October 15 with the R-C spectrograph on the CTIO 4m
telescope. It is shown in Fig.~2 (reproduced from Peroux et al 2000,
in prep). The spectrum is characterized by emission lines at
$\sim$6600\AA\ from the Lyman-$\alpha$(1216\AA)+NV(1240\AA) blend and
CIV(1549) at $\sim$8300\AA.  The Lyman-$\alpha$ line has a peak at
6587$\pm$10\AA\  corresponding to a redshift of 4.418. The observed
equivalent width of the Ly-$\alpha$/NV blend is 191\AA\
(EW(rest)=35\AA). This is relatively small compared with radio--quiet
$z>$4 quasars, but similar to that of the other strongly X-ray
emitting radio--loud quasars (see Figure 4 in Hook \& McMahon 1998).
The peak of the CIV emission line is at 8340$\pm$10\AA\  corresponding
to a redshift of 4.384. The observed and rest frame equivalent widths
are 131\AA\ and 30\AA, respectively.  The CIV line is asymmetric with
a strong CIV absorption doublet (observed EW=13.5\AA) at
$z=4.4325\pm$0.0005 i.e. redshifted with respect to the peak of the
CIV emission line by 2700 km s$^{-1}$.  This absorption system also
displays absorption from SiIV(1394,1403) and NV and Lyman-$\alpha$
where the observed EW of Lyman-$\alpha$ is 7.9 (rest=1.45) \AA.

There is no evidence for an optically thick Lyman-limit system at the
redshift of the quasar and we set a conservative upper limit of the
$N(HI)$ column density at $< 3\times 10^{17}$ cm$^{-2}$
i.e. $\tau$$<$1.  To be consistent with the above Lyman-$\alpha$
absorption line strength and the constraint on $N(HI)$, the doppler
parameter $b>$40 km s$^{-1}$.

The emission line redshift difference of 1800 km s$^{-1}$ is smaller
than the CIV line width (FWHM) of 11,500 km s$^{-1}$ and is within the
range observed in other $z>$4 quasars (1100$\pm$1790 km s$^{-1}$;
Storrie-Lombardi et al 1996) and in lower redshift quasars (Tytler \&
Fan 1992). We take the redshift of the quasar to be the mean of these
two lines at $z=4.401\pm$0.006. CIV absorption within $\pm$5000 km
s$^{-1}$ of the CIV emission line redshift is quite common in
radio--loud quasars, though its origin is debatable (Anderson et al
1987).

The continuum flux at a rest frame wavelength of 1450\AA\ is 
5.8$\times$ 10$^{-17}$ erg s$^{-1}$ cm$^{-2}$ \AA$^{-1}$, while 
the spectral index is difficult to measure due to the presence
of broad weaker emission lines that lie between NV and CIV.


\begin{figure}
\psfig{file=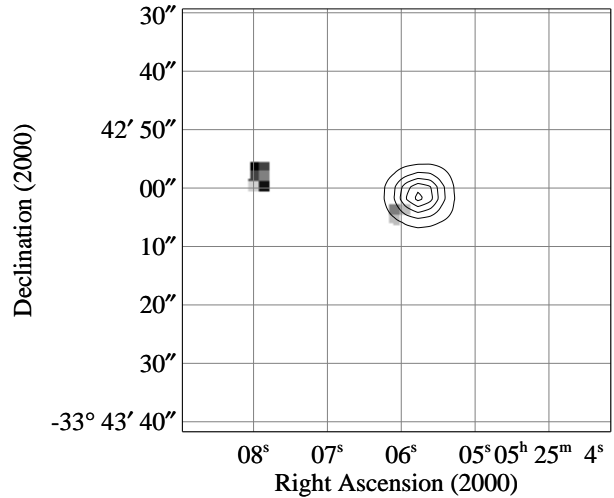,width=8truecm}
\caption{Overlay of the optical (POSS) and the ROSAT HRI contours of
PMN~J0525-3343. The radio position coincides with the faint optical
source (greyscale) at the center of the plot.}
\end{figure}

\begin{figure}
\psfig{file=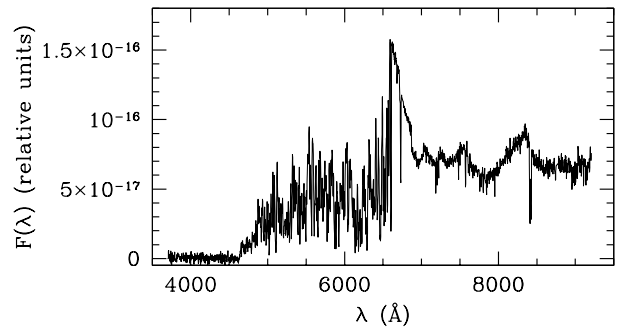,width=8truecm}
\caption{Optical spectrum; the Lyman $\alpha$ and CIV broad emission
lines are clearly visible.}
\end{figure}

\section{Radio data}

PMN~J0525-3343 was observed for two hours on 1998 December 30 with the
Very Large Array in the C (3 km) configuration, and for 2 hr on 1999
February 26 in the D (1 km) configuration.  The radio source was
observed at 8.4 GHz, 22 GHz, and 43 GHz. The spatial resolutions for
the 3 km array observations were: $7.9''$x$2.4''$, $2.5''$x$0.86''$,
and $1.2''$x$0.4''$, respectively (FWHM of major and minor axes of the
Gaussian restoring beam, with major axis position angle oriented
north-south at all frequencies). The absolute flux scale was set by
observing 3C~286.  The radio source position was found to be: 05 25
06.18$\pm$0.16, -33 43 05.55 $\pm$ $0.05''$ (J2000).  Its flux density
on 1998 December 30 was 41.0 $\pm$ 0.8 mJy at 8 GHz, 12.1 $\pm$ 0.5
mJy at 22 GHz, and 5.5 $\pm$ 0.6 mJy at 43 GHz. On 1999 February 26 it
was 39.9 $\pm$ 0.8 at 8 GHz, 11.1 $\pm$ 0.7 at 22 GHz, and 6.5 $\pm$
0.9 at 43 GHz.  Thus no significant variability has been detected on a
time scale of two months, with limits of $\le$ 3 per cent at 8 GHz, 7
per cent at 22 GHz, and 18 per cent at 43 GHz.  The source appeared
spatially unresolved at all frequencies.  The 43 GHz image constrains
the size to be less than 0.4$''$ in the east-west direction, and less
than about 1$''$ in the north-south direction. 

The spectral index over the observed frequency range of 8 to 43 GHz is
$\alpha_r$=1.2. The steep spectral index for this blazar may be due to
the high rest frame frequency range of 44 GHz to 240 GHz. Steep
spectrum cores at observed frequencies $\ge$ 8 GHz have been seen in
many $z >$ 2 radio--loud quasars and radio galaxies (Lonsdale, Barthel
\& Miley 1993, Carilli et al 1997). Lonsdale et al (1993) present
models in which the dominant component in the core-jet may not be
synchrotron self-absorbed at these high frequencies, thereby setting a
lower limit of about 1 mas to the typical size of the dominant core
radio component. 

Our present observations do not constrain the sub-arcsec radio
structure of the object; very long baseline interferometry is in
progress.

\section{X--ray data}

The ROSAT HRI count rate of the source on 1998 March 4--6 was
$0.012\pm0.001$ ct~s$^{-1}$. PMN~J0525-3343 was next  observed with
ASCA on 1999 March 1-2. The data reduction was done using FTOOLS
version 4.2 and the standard calibration provided by the ASCA Guest
Observer Facility (ASCA GOF) at NASA/Goddard Space Flight Center. Good
exposure times for each detector pair are 34 ks for the SIS and 31 ks
for the GIS.

We consider the 0.8--10 keV data from both the SIS and GIS detectors,
taking into account the possible systematic error discussed below.
Although a significant deficiency in the SIS detector response curve
below 1 keV has been reported for observations carried out after 1994
(ASCA GOF 1999), which could lead to an overestimate in measured
absorption column densities, this effect should not be relevant here
since the statistical error dominates over the systematic one in this
faint X-ray source. In the energy range we use, the SIS and GIS data
are indeed in good agreement and therefore an X-ray deficit in the
low energy band, as reported below, is not an artifact of the
calibration uncertainty. The SIS data in the 0.6--0.8~keV band are
consistent with the trend found from the higher energy data. The
result is also confirmed by the $Beppo$SAX/LECS data.

A fit with a simple power-law modified only by the Galactic absorption
(a column density of $2.2\times 10^{20}$ cm$^{-2}$; Elvis, Lockman \&
Fassnacht 1994) yields a photon index ($\Gamma=\alpha+1$) of
$\Gamma=1.38\pm 0.06$ with $\chi^2=201.0$ for 210 degrees of freedom
(uncertainties on all spectral parameters are quoted at the 90 per
cent confidence level). This fit is statistically acceptable, but
leaves systematic, curved residuals across the fitted energy range.
This spectral feature can be accounted for by introducing a flattening
in the low energy part of the spectrum either due to excess absorption
-- local or intrinsic -- or to an intrinsic change in the (power--law)
slope. The inclusion of the low energy spectral flattening improves
the quality of the fit by about 10 in $\chi^2$ for both models (see
Table~1; these improvements are significant at the 99.9 and 99 per
cent confidence levels for the absorption and broken power-law models,
respectively). We assume neutral absorption and solar abundances
(solar or higher nuclear abundances appear to be present even at such
high $z$, eg Shields
\& Hamann 1997).

$Beppo$SAX observed the source on 2000 February 27-28, with the three
main detectors of the Narrow Field Instrument package. The exposure
times for each detector are 24 ks, 60 ks and 29 ks for the LECS, MECS
and PDS, respectively.  The background subtracted PDS count rate is
$0.054\pm 0.024$ cts s$^{-1}$, which is just above the systematic
error of the instrument ($\sim 0.03$ cts s$^{-1}$, Guainazzi \&
Matteuzzi 1997). We therefore consider the PDS detection as not
significant. The extrapolation of the best-fit model obtained from the
LECS and MECS data is consistent with the PDS result (marginally
significant detection would be achieved at the ASCA flux
level). Spectral flattening is evident at low energies; $\chi^2$ drops
by 2 for intrinsic absorption and 7 for a broken power law.

\begin{table*}
\caption{{\bf Results of the spectral fits.} 
Confidence ranges are 90 per cent for one parameter.}
\vskip 0.2 true cm
\begin{tabular}{l c c c c c c}
\hline \hline Model & $N_{\rm H}$ & $\Gamma_1$ & $E_{\rm break,obs}$ &
$\Gamma_2$ & $\chi^2$/d.o.f. \\

       & (10$^{21}$ cm$^{-2}$) & & (keV) & \\

\hline 

ASCA &&&&& \\

Power-law & $N_{\rm H,gal}$ & 1.38 (1.32-1.44) & && 201.0/210 \\

Broken power-law & $N_{\rm H,gal}$ &0.93 (0.52-1.15) & 1.72 (1.35-2.12) &
1.54(1.43-1.66) & 189.7/208\\

Power-law + free local $N_{\rm H}$ & 2.0(1.1-2.9) & 1.60(1.47-1.73) & -- &
-- & 190.0/209\\

Power-law + intrinsic $N_{\rm H}$ & 180(68-260) & 1.67(1.51-1.85)& --
& -- & 190.1/209 \\

\hline 

$Beppo$SAX+ASCA &&&&& \\

Power-law & $N_{\rm H,gal}$ & 1.40 (1.34-1.46) & && 233.2/237 \\

Broken power-law & $N_{\rm H,gal}$& 0.90(0.54-1.12)&  1.72(1.35-2.10)
& 1.57(1.47-1.68) &216.3/235\\

Power-law + free local $N_{\rm H}$ & 1.9(1.1-2.8) & 1.63(1.52-1.75) & -- &
-- &  216.7/236\\

Power-law + intrinsic $N_{\rm H}$ & 170(95-360) & 1.67(1.54-1.82) & -- &
-- & 218.5/236 \\

\hline \hline
\end {tabular}
\end{table*}

As the fit to the $Beppo$SAX data is consistent with the ASCA one (no
indication of spectral changes between the two observations can be
found) we performed a joint fit of the two datasets, leaving the
relative normalization free.  The flattening is convincingly seen from
both instruments (see Fig.~3) and both the (intrinsically) absorbed
power-law model and the broken power--law give statistically good fits
to the combined data (see Fig.~4, Table 1).

The observed flux is $F{\rm (2-10\ keV)}=1.2\times 10^{-12}$ erg cm$^{-1}$
s$^{-1}$ (ASCA).  No statistically significant variations were found
within the individual observations -- weak variability is only hinted
at in the second half of the ASCA observation, with an amplitude not
exceeding 30 per cent (see Fig.~5).  However the flux change between
the ASCA and $Beppo$SAX datasets corresponds to a variation of 60 per
cent over $\sim$ 2 months in the source frame, i.e. the flux ratio
ASCA/$Beppo$SAX is 1.6$\pm$0.2. Note that the agreement in flux
calibration between $Beppo$SAX (MECS) and ASCA is better than 3 per
cent, according to the results of the (quasi-) simultaneous
observations of 3C273 (see Yaqoob \& ASCA Team 2000).  Although this
intercalibration might depend on the spectral shape of a source, the
spectra of 3C273 and PMN~J0525-3343 are similarly hard and thus we
expect that the agreement should be at the same level.  Therefore
systematic errors between the two instruments are much smaller than
the statistical one.

Only marginal variability has been instead found between the ASCA and
$Beppo$SAX observations and the ROSAT HRI one: the HRI count rate in
the 0.1--2.4 keV band corresponds to a flux of $4.20\times 10^{-13}$
erg cm$^{-2}$ s$^{-1}$, assuming Galactic absorption and the
ASCA/$Beppo$SAX photon index of $\Gamma=0.9$, to be compared to $5.2
(3.3)\times 10^{-13}$ erg cm$^{-2}$ s$^{-1}$ for the ASCA ($Beppo$SAX)
observations, respectively.

\begin{figure}
\psfig{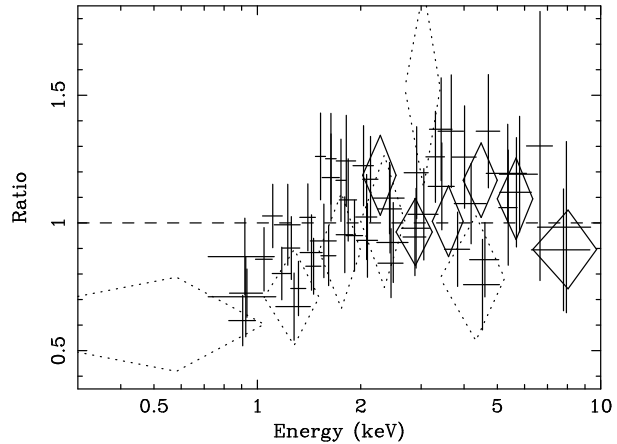}
\caption{Data to model ratio for the ASCA and $Beppo$SAX data fitted
above 2 keV. The diamonds refer to the $Beppo$SAX datasets (solid
diamonds: MECS, dashed diamonds: LECS). Data from all the four ASCA
detectors are plotted.}
\end{figure}

\begin{figure}
\psfig{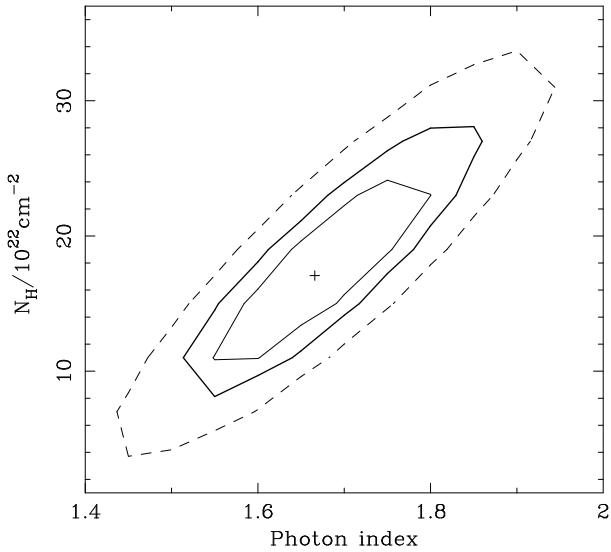}
\caption{Confidence contours (68, 90 and 99 per cent) for the excess column
density and photon index from joint fitting of the ASCA and $Beppo$SAX
data.}
\end{figure}

\begin{figure}
\psfig{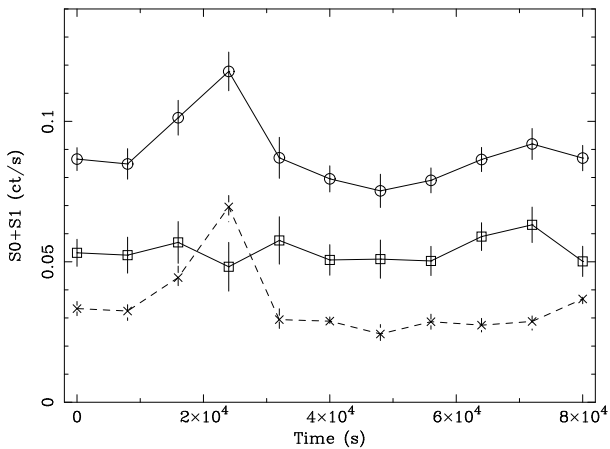}
\caption{ASCA SIS light curve: the source extracted region (circles),
a source-free region of the detector with similar area (crosses),
background-subtracted light curve (squares).  }
\end{figure}

The integrated hard X--ray luminosity, not corrected for absorption,
is $L{\rm (2-10\ keV,\ ASCA)}=5.0\times 10^{46}$ erg s$^{-1}$ (rest frame
range $\sim$ 11 -- 54 keV).

Finally, we note that the observed X--ray variability, which sets
limits on the emitting region size $< 3\times 10^{17} \delta_1$ cm,
$\delta_1=\delta/10$ being the Doppler factor, and the broad band
spectral indices (i.e, $\alpha_{ro} = 0.81, \alpha_{rx} = 1.03,
\alpha_{ox} = 1.22 $, where the optical radiation is assumed unbeamed)
suggest that the emission properties of PMN~J0525-3433 resemble those
of blazars (see also Fabian et al 1999).

\section{Discussion}

PMN~J0525-3433 (here), GB~1428+4217 (Boller et al 2000) and RXJ
1028.6-0844 (Yuan et al 2000) are the most distant objects for which
it has been possible to determine X-ray spectral features, i.e. a
flattening of the lower energy X--ray spectrum.  Intriguingly, the
inferred soft energy (observer frame) spectral slope of PMN~J0525-3433
follows the trend found by Fiore et al (1998) and Reeves \& Turner
(2000) for sources up to $z\sim$ 4.2.  As it is not yet clear whether
this is due to external or intrinsic absorption, or to the shape of
the emitting particle or soft photon spectra we will discuss all
possibilities in turn.

\subsection{Absorption}

The first issue concerns whether absorbing material is located along
the line of sight or is intrinsically associated with the system. The
hypothesis in which the spectral flattening is due to changes in the
source, either evolutionary or cosmological changes, is favoured by
the fact i) that the flattening seems to be associated mostly with
higher redshift radio--loud quasars (although the data are sparse, the
only two $z>1$ radio-quiet quasars showing a flattening in the Reeves
\& Turner sample seem to be atypical sources); ii) that as well as
PMN~J0525-3433 the other three $z>4$ quasars show this feature: it
therefore appears unlikely that an (unusually thick) absorber is
located along the lines of sight to all of them.


It is therefore interesting to explore viable origins of the putative
intrinsic absorbing gas.  Note that the results on PMN~J0525-3343 seem
to definitely exclude that the trend in flattening is due to a shift
of a soft X--ray excess component out of the observational band in the
higher redshift objects: since the soft X-ray spectrum is {\it
flatter} than, and not similar to, the hard X--ray one. \footnote{Note
also that the ROSAT PSPC (data used by Fiore et al) would not have
enabled a flattening in PMN~J0525-3343 (and thus in all lower redshift
objects) to be seen, as the break energy would have been close to the
upper extreme of its spectral coverage.}

It is possible that the gas occupies a large scale, although this
requires high values of the mass of the absorber. It might, for
example, be associated with a galactic outflow, perhaps following the
formation and initial activity of the central black hole, which expels
gas from the host--galaxy (Silk \& Rees 1998, Fabian 1999), with
inferred mass outflows of the order of $\dot M \sim$ few$\times 10^3
M_{\odot}$ yr$^{-1}$.

On the other hand, Elvis et al (1998) have pointed out that in the
`flattening' sources the presence of absorbing line systems at high
velocities (up to 10,000 km s$^{-1}$) and/or reddening, are unusually
common. Thus a nuclear origin is favoured, at least for the
highest line velocity systems.

\subsubsection{X-ray vs optical-UV absorption}

We stress however that serious problems arise, independently of the
location or nature of the absorbing material.  The column densities
inferred from the X--ray spectrum imply, for a standard gas-to-dust
ratio and composition, an extinction in the optical--UV band far in
excess of that allowed from the detected optical fluxes. The UV
reddening constraints indicate that the column density in neutral
hydrogen is less than $3\times 10^{17}$\psqcm\ to the nucleus whereas
the X-ray spectrum requires a column corresponding to $\sim 2\times
10^{23}$\psqcm (equivalent to $A_V \sim 100$).

High metal enrichment or the presence of large dust grains (presumably
not yet fragmented at high redshift) are possibilities which can
account for the qualitatively similar discrepancy between the inferred
$N_H$ and $A_V$ in both local radio--quiet sources (e.g. Salvati \&
Maiolino 2000) and in radio--galaxies (e.g. Cen A). Nevertheless, such
possibilities do not seem sufficient to account for the large
discrepancy found in PMN~J0525-3343.

A possibility is that the optical photons are scattered along the
line of sight. Strong (if unbeamed) emission and a high covering
factor for the scattering medium are required. Note that the optical
component cannot be totally dominated by beamed emission, although
the intrinsic EW of the lines appear rather low compared to those of
unbeamed objects. Although possible for one object it would be very
unlikely if this geometry applied to all three objects yet studied
above $z>4$.

One alternative solution to the millionfold discrepancy is that the
absorbing gas is highly ionized and dust destroyed, because of the
nuclear source itself or high temperature, so that the absorber is,
say, OVII (most absorption will be due to oxygen at the observed
energies).
The absorber therefore resembles a warm absorber such as is common in
Seyfert galaxies (Reynolds 1997).
Indeed, the presence of absorption features associated with CIV (and
Lyman-$\alpha$ + NV), which are seen in almost all luminous quasars
with X--ray absorption column densities comparable to that inferred
from PMN~J0525-3343 (e.g. Brandt, Laor \& Wills 2000), supports this
hypothesis. (The CIV and X-ray absorbers in low $z$ objects are not
due to exactly the same material, but are clearly associated in some
way.)


In order to verify the viability of the warm absorber hypothesis, we
have performed X--ray spectral fits using the results from the
photoionization code CLOUDY, allowing the column density and
ionization parameter $\xi = L / n R^2$ to vary. $L, n$ and $R$ are the
luminosity of the ionizing source, the density of the absorber and its
distance from the source, respectively. The intrinsic spectral index
of the ionizing radiation is fixed at the value $\Gamma = 1.75$
obtained from the fitting of the highest X-ray energy spectrum.  The
results are shown as confidence contours of $\xi$ vs $N_H$ in Fig.~6,
and are well consistent with a moderate ionization parameter of $\xi
\sim 10$ and a high column $N_H \sim 3\times 10^{23}$
cm$^{-2}$. The attenuation of the optical--UV spectrum obtained is
consistent with observations, provided that $\xi > 10$ (so that the HI
photoelectric edge remains small; i.e. $\tau<1$).


\begin{figure}
\psfig{file=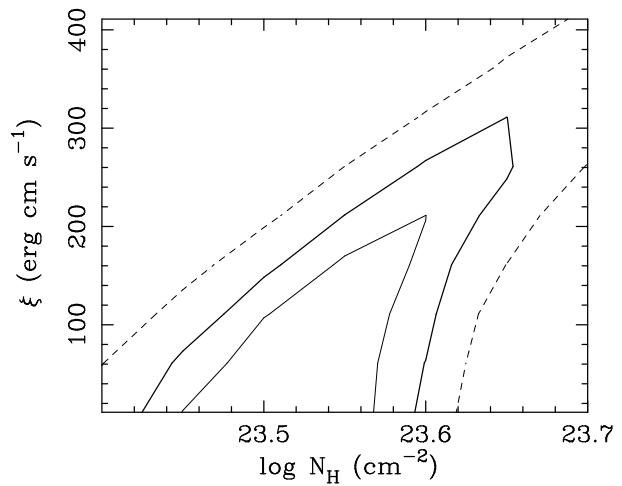,width=8truecm,angle=270}
\caption{Contour plot of the ionization parameter $\xi$ vs $N_H$ 
(68, 90, 99 per cent levels for two parameters).}
\end{figure}

\subsection{Intrinsic spectral properties}

An alternative view might invoke evolution with redshift of the
spectral characteristics of either the emitting particle distribution
and/or an external photon field. The most accepted scenario for blazar
emission attributes the production of the high energy component to
inverse Compton scattering of soft photons on relativistic electrons.
For the powerful sources an intense quasi--isotropic radiation field,
peaked at frequency $\nu_{\rm ext}$, is believed to be produced in the
nuclear regions. This dominates over the synchrotron photon energy
density emitted locally inside the jet, and provides the seed photons
for scattering (Sikora, Begelman \& Rees 1994). This scenario has been
quite convincingly tested by fitting the spectral energy distributions
of tens of local ($z< 2$) blazars (e.g. Ghisellini et al 1998) and at
least two at the largest redshifts: GB~1428+4217 at $z=4.72$ (Fabian
et al 1998) and in Fig.~7 we show the specific modeling for
PMN~J0525-3343 (see the figure caption for the parameters involved in
the fit).  Note in particular that there is an apparent tendency for
the high redshift sources to require more extreme values of the
intensity of the external soft photon field compared with lower
redshift objects, in turn leading to larger cooling through scattering
on external photons.

Note that the spectral break observed here is significantly different
from that already found in some blazars, where both convex and concave
(e.g. Wolter et al 1998) X--ray spectra are observed. The former ones
are interpreted as due to the contribution of both the steep
synchrotron and flat inverse Compton emission, at lower and higher
frequencies, respectively, while the latter ones are well reproduced
as the tail of the synchrotron component: there however the higher
energy part of the spectrum is extremely steep, at odds with the
flat high energy component detected in PMN~J0525-3343.

\subsubsection{A cutoff in the particle distribution}

A flattening of the spectrum could indeed be expected if there is a
low energy cutoff in the emitting particle distribution.  In fact,
within the above picture, a low energy cutoff in the relativistic
particle distribution at $\gamma_{\rm min}$ would produce a flattening
in the scattered spectrum below $\sim \gamma_{\rm min}^2 \Gamma^2
\nu_{\rm ext}$, where $\Gamma$ is the bulk Lorentz factor. The
flattest slope which could in principle be obtained corresponds to the
energy distribution produced by the scattering of a single electron on
an isotropic field, i.e. $\alpha=-1$.  Therefore the results on
PMN~J0525-3433 are fully consistent with such a hypothesis
\footnote{Note that also the findings on RXJ1028.6-0844 (Yuan et al
2000) are consistent with a flattening of the spectrum rather than
excess absorption.}.  In particular if $\nu_{\rm ext} \sim 10^{15}$ Hz
(rest frame), the break observed in the X--ray spectrum of this source
would imply a cutoff in the particle distribution at a Lorentz factor
$\gamma_{\rm min} \sim$ a few.

The importance of such a finding would be two-fold. On one side
$\gamma_{\rm min}$ constitutes a crucial parameter to determine the
nature of the power and matter flowing in the jet, as it determines
the total density of (relativistic) particles present in the flow
(Celotti \& Fabian 1993; Sikora \& Madejski 2000). On the other side,
a cutoff in the particle distribution implies, as the cooling
timescales are extremely short, that an highly efficient
re--acceleration process is operating at the same rate.

Problems arise also with this interpretation (see also Fiore et
al 1998), as there is no evidence of intrinsic breaks in the X--ray
spectra of nearby radio--loud quasars which would be well detectable
if located at the corresponding intrinsic energy $\sim$ 8 keV. Indeed
one might expect that in less powerful (nearby) objects, the density
in the external radiation field would be lower and the cooling less
dramatic and thus under these conditions the cooling break in the
particle distribution would be located at even higher energies.

\subsubsection{Sharply peaked soft photon distribution}

An alternative scenario occurs if the seed photon distribution is
peaked, or at least rapidly decreases below $\nu_{\rm ext}$.  If so,
the distribution of the scattered photons (by the lowest energy
electrons) would reproduce such a slope (e.g. Ghisellini 1996). As the
scattering electrons are moving with Lorentz factor $\Gamma$, in the
observed frame the scattered radiation will appear boosted in
frequency by a factor $\sim \Gamma^2$. Thus below $\sim \Gamma^2
\nu_{\rm ext}$ a spectral decline will be observable. The sharpness of
such a spectral break could be accounted for by the strong anisotropy
of the scattered radiation (we see only photons within a small angle).

Clearly such a possibility is highly speculative given the lack of
information on the nuclear soft photon field distribution. Note
however, that if the electron distribution extends down to low
energies, the location of the X-ray spectral break in PMN~J0525-3343
would indicate the beaming Lorentz factor of the emitting plasma,
which then must be $\sim 40$, if $\nu_{\rm ext} \simeq 10^{15}$ Hz.

The main difficulty arising with scenarios attributing the flattening
to intrinsic properties of the electron/soft seed photon
distributions, is to convincingly account for the intriguing trend
with $z$ of the soft X--ray spectrum of radio--loud objects which
systematically flatten with redshift (Fiore et al 1998).  An
interesting possibility, within the external Compton scenario, is that
by increasing the redshift (and luminosity) of the source, the ratio
of external Compton (EC) to synchrotron self--Compton (SSC) increases.
If so, in high power sources the EC strongly dominates the emission,
and leads to the depression (extreme flattening of soft X--rays) in
the spectrum, while in increasingly lower redshift (weaker) sources
the SSC contributes to the soft-medium X--ray emissions, filling the
spectral depression (e.g. Sikora et al 1994). Clearly such an increase
in the external photon energy density could be associated both with
the source power and with the nuclear conditions of quasars at higher
$z$.

\begin{figure}
\psfig{file=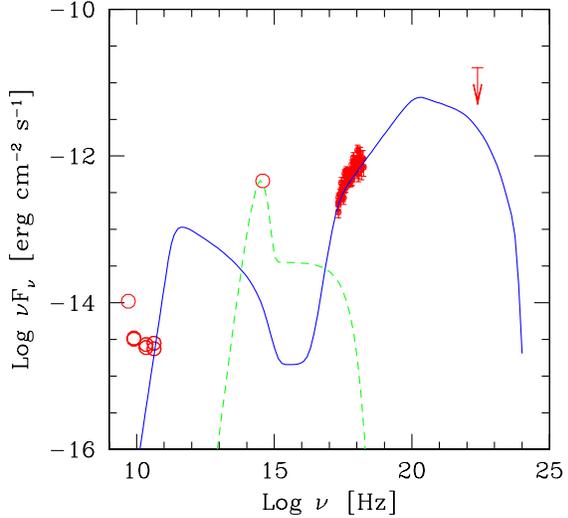,width=8truecm}
\caption{Broad band Spectral Energy Distribution. The data are from
this work, except the EGRET upper limit. The curves represent a model
assuming a homogeneous region of a relativistic jet, predominantly
emitting via synchrotron and Compton scattering an external soft
photon field (whose distribution is shown by the dashed line).  For
details see e.g. Ghisellini et al (1998). The model parameters are the
following: size of the region $\sim 5\times 10^{16}$ cm, compactness
of the injected particle luminosity $\ell_h \sim$ 0.1, compactness of
the external radiation field $\ell_s \sim 3$, magnetic field $B \sim
3$ G.}
\end{figure}

\section{Conclusions}

We have studied the X-ray spectral properties of the high redshift
radio--loud quasar PMN~J0525-3343, by means of ROSAT, ASCA and
$Beppo$SAX observations. We have found that the spectrum flattens at
the lowest energies, indicating either a spectral flattening or a
significant absorption in excess of the Galactic one.  This finding is
consistent with and extends the range in redshift of the systematic
trend of increasing flattening with increasing redshift found
previously for radio--loud sources. Several possibilities have been
discussed.

The most plausible hypotheses include either an intrinsic flattening
of the soft X--ray spectrum, originating from the shape of the energy
distribution of particles and/or seed photons involved in the
production of the Compton component, or the presence of warm absorbing
gas in the nuclear region. If this is similar to the absorber found in
local (Seyfert) objects, its distance from the central source would be
of order $\sim 10^{16}$ cm. Evidence for CIV absorption in the
spectrum supports the warm absorber hypothesis. The equivalent width
of the absorption and value of $\alpha_{\rm ox}$ for PMN J0525-3343
fall on the correlation of Brandt et al (2000; Figure 4). Indeed the
objects neighbouring PMN J0525-3343 in that correlation are observed
to have X-ray warm absorbers. The velocity difference between the CIV
absorption and emission are unlikely, in an object at $z=4.4,$ to be
due to an intervening galaxy in a massive cluster of galaxies, since
such clusters would not have then formed. Possibly the absorption line
redshift is that of the quasar nucleus, and the broad emission line
profiles have been altered by further, unresolved, absorption into
peaking at a lower redshift.   

In the former case, the column-redshift dependence could be accounted
for by the decreasing contribution of synchrotron self--Compton
emission at the softest energies due to an increasingly intense
external photon field. In the latter, one might envisage an increase
in the amount of nuclear gas and/or of ionizing radiation with
$z$. Interestingly, it has been suggested (e.g. Reeves \& Turner 2000
and references therein) from the observed characteristics of the
spectra of radio--quiet sources, that the key parameter regulating
their spectral properties is the mass accretion rate in Eddington
units: high luminosity sources would thus be characterized by an
increasing ionization state of the reflecting disc and possibly of any
nuclear absorber. A similar trend might be at the origin of the
increasing photon density required by the above scenario in
radio--loud objects.

Future observations with Chandra and XMM-Newton, and in the optical,
will enable us to discriminate between the models for these sources.

Clearly, this issue needs more investigation, since it represents a
potential means for understanding the conditions and evolution of the
nuclear/galactic environment and the galaxy--quasar connection.  In
particular future X--ray observations with wide energy coverage and
good spectral resolution as provided by a significant number of high
redshift radio--loud objects, will allow us to a) establish the origin
of the spectral flattening, determining the abundance and ionization
state of any absorber and -- if variability or clear features are
detected -- its spatial location and redshift/velocity and b) assess
its evolutionary properties. Information from the optical--UV spectrum
are also of prime importance, as the amount of (or limit on) the
extinction can be directly compared to the X--ray one and any line
absorption feature would determine the absorbing gas velocity,
although photometry alone might not be sufficient to distinguish
between the steep non--thermal continuum typical of blazars and a
reddened quasar spectrum.

\section*{Acknowledgments}

The VLA is operated by the National Radio Astronomy Observatory, which
is a facility of the National Science Foundation operated under
cooperative agreement by Associated Universities, Inc. CLC would like
to thank G. Taylor for useful discussions. The Royal Society (ACF),
the Italian MURST (AC) and the NASA LTSA grant NAG 5-8107 (WNB) are
acknowledged for financial support.

\end{document}